\documentclass[aps,prd,amsfonts,twocolumn,groupedaddress,nofootinbib]{revtex4}

\newcommand{\be}{\begin{equation}}
\newcommand{\ee}{\end{equation}}
\newcommand{\ba}{\begin{eqnarray}}
\newcommand{\ea}{\end{eqnarray}}

\usepackage{graphics,epsfig}
\usepackage{epstopdf}

\usepackage{color}

\begin{document}

\preprint{WM-05-101}

\title{Beam Single-Spin Asymmetry in Semi-Inclusive Deep Inelastic Scattering}

\author{Andrei V. Afanasev$^{(a,b)}$ and Carl E. Carlson$^{(c)}$}

\affiliation{
$^{(a)}$Thomas Jefferson National Accelerator Facility, Newport News, VA 23606, USA\\
$^{(b)}$Department of Physics, Hampton University, Hampton, VA 23668\footnote{Present address}\\
$^{(c)}$Department of Physics, College of William and Mary, Williamsburg, VA 23187, USA }

\date{November 29, 2006}

\begin{abstract}

We calculate, in a model, the beam spin asymmetry in semi-inclusive jet production in deep inelastic scattering.  This twist-3, $T$-odd observable is non-zero due to final state strong interactions.  With reasonable choices for the parameters, one finds an asymmetry of several percent, about the size seen experimentally.  We present the result both as an explicit asymmetry calculation and as a model calculation of the new transverse-momentum dependent distribution function $g^\perp$.
\end{abstract}

\maketitle



\section{Introduction}


Semi-inclusive deep inelastic scattering (SIDIS) provides a way to reach a more detailed understanding of the structure of a hadronic target~\cite{baronereview,jj,mt}.  The information gained can be codified into transverse momentum dependent distribution functions (TMD's), which in leading twist are related to quark probability distributions.  Single spin asymmetries provide a method to isolate and measure specific TMD's~\cite{leveltmulders,Yuan:2003gu}, often those of higher twist.  Experimentally, both beam and target spin asymmetries have been observed~\cite{hermes_ssa,Avakian:2003pk}.

This paper is concerned with calculations of beam single spin asymmetries in semi-inclusive production of a quark jet.  A beam spin asymmetry can be viewed in a number of ways.  At the simplest level, finding a beam spin asymmetry means defining ``up'' and ``down,'' say by taking a cross product of the incoming electron spin $\vec S_e$ with the virtual photon three-momentum $\vec q$,  and then seeing if the outgoing hadron or jet is more likely to emerge with transverse momentum up or down relative to the electron scattering plane as just defined.   Going further, a jet is a collimated spray of hadrons,  where all the final-hadrons in the jet are summed and their phase space integrated, and we represent the jet momentum by the quark momentum $\vec p_q$.   A more formal observable that corresponds to the beam spin asymmetry is 
$\vec S_e \times \vec q \cdot \vec p_q$.  Non-zero beam spin asymmetry means a non-zero average, weighted by the cross section, value of this observable.   One can define an azimuthal angle $\phi$ as the angle between the electron scattering plane and the plane defined by the photon and quark momenta.  The observable has a $\sin\phi$ dependence, so that a non-zero beam spin asymmetry requires a term in the differential cross section proportional to $\sin\phi$, and the relative size of this term is the measure of the beam spin asymmetry that we shall use in the body of this paper.\footnote{If parity is violated, as in neutrino scattering, there is an additional term in the single-spin asymmetry proportional to $\sin 2 \phi$}

The electron interacts with the hadrons by photon exchange, so one may ask how the photon obtains the spin information and carries it across.  The detail is that longitudinally polarized electrons, which we consider here, give rise to photons that have a vector polarization $\vec S_\gamma$, and that this vector polarization is not parallel to the photon momentum although it is in the electron scattering plane.  (Virtual photons emitted by unpolarized electrons have only tensor polarization.)  An alternative beam spin asymmetry observable is $\vec S_\gamma \times \vec q \cdot \vec p_q$.

The observable $\vec S_e \times \vec q \cdot \vec p_q$ is odd under time-reversal ($T$-odd).  Hence the beam spin asymmetry is and must be zero in a lowest order calculation.  However, it is not zero in higher orders, as hadronic final state interactions lead to a relative phase between the longitudinal and transverse amplitudes, and it is interference between these two amplitudes gives the beam spin asymmetry.

Single spin asymmetries in semi-inclusive reactions have their own history.  Part of the history is common to beam and target single spin asymmetries, since both are $T$-odd, though it developed first in the target spin asymmetry context.  Early on Sivers~\cite{sivers91} pointed out that target spin asymmetry could be used to measure a new distribution function for the proton.  However, it was quickly noticed that the asymmetry was $T$-odd and so it was first thought that the asymmetry had to be zero~\cite{collins92}.  So it remained until Brodsky, Hwang, and Schmidt~\cite{bhs} demonstrated explicitly in a model the statement already made above, that because of final state interactions the $T$-odd operators could give experimentally non-zero expectation values.  Shortly thereafter, this statement was confirmed from other viewpoints~\cite{collins,Belitsky:2002sm}.  The same simple model has since been applied to a number of other target spin dependent TMD's~ \cite{Gamberg:2003pz}.

The other part of the history impacted upon questions of factorization in SIDIS.  It is believed or hoped---and perhaps proofs are becoming available~\cite{Ji:2004wu,Collins:2004nx} for leading twist---that cross sections for SIDIS can be written as a sum of terms, where each term is a convolution of a distribution function that depends only on the target and a fragmentation function that depends only on the final state.  For jets observed in the final state, the fragmentation function would be either absent or trivial.  However, until relatively recently, the lists of possible target TMD's contained none that could give a beam spin asymmetry when the observed hadron was the whole quark jet~\cite{mt,efremovgoeke}.  A problem arose when we, in an earlier note~\cite{Afanasev:2003ze}, and Metz and Schlegel~\cite{Metz:2004je} directly calculated the beam spin asymmetry and found a non-zero result without using any $T$-odd fragmentation function.  One could then speculate that factorization did not work for this situation since there was then no parton distribution associated with this asymmetry.  However,  Bacchetta {\it et al.}~\cite{Bacchetta:2004zf} and Goeke {\it et al.}~\cite{Goeke:2005hb}, have made progress by uncovering an initial state distribution function $g^\perp$, which was originally dismissed because it was $T$-odd, and which precisely gives a beam spin asymmetry with an outgoing jet.  Factorization for observable
matching $g^\perp$, and twist-3 distributions, has yet to be proved. (We might note that if specific hadrons rather than a jet are observed, there are and were known mechanisms to obtain a beam spin asymmetry~\cite{leveltmulders,Yuan:2003gu, Efremov:2002ut,Wakamatsu:2003uu}.  Also, there were early perturbative QCD calculations including gluon loop diagrams which showed beam asymmetries of the order of a percent~\cite{Hagiwara:1982cq}.)

The model we use~\cite{bhs,Afanasev:2003ze,Metz:2004je,ezawa-drell-lee} is a simple one where the proton is represented as a bound state of a quark and a scalar diquark, and where the hadronic final state interaction is produced by a gluon exchange.  The beam spin asymmetry involves both longitudinal and transverse photon polarizations.  The longitudinal matrix element (for spin-1/2 quarks) is subleading for large $Q$.  For reasons of electromagnetic gauge invariance, we keep all Feynman diagrams where the photon attaches to a charged particle.  Hence we have three diagrams, in general, in lowest order, and there are also three diagrams in one-loop order that have an imaginary part.  The diagrams are shown below in Figs.~\ref{fig:lowest} and~\ref{fig:oneloop}.  For calculations where only the transverse amplitudes are needed to leading order in $1/Q$, such as the unpolarized cross section or the target spin asymmetry~\cite{bhs}, only the diagrams where the photon interacts directly with the quark are needed.

We show our calculations in Sec.~\ref{sec:calc}, with the results for the beam spin asymmetry given both as explicit formulas, for the limit that the masses and jet transverse momentum are all smaller than the invariant momentum transfer $Q$, and as plots for selected values of the parameters.   
Our calculation of the beam single spin asymmetry does not depend on factorization theorems. Now, with modern knowledge, finding a beam spin asymmetry does not disprove factorization, and if factorization is correct for the situation at hand, we can interpret the beam spin asymmetry in terms of the distribution function $g^\perp$.  We do so in Sec.~\ref{sec:gperp}, and include a few further comments about factorization. 
The beam spin asymmetry we discuss here has no mention of hadronic spin, so one expects we could exhibit a similar asymmetry in a model with scalar quarks.  We remark further about this in Sec.~\ref{sec:scalar}.  Finally, we end with a summary in Sec.~\ref{sec:end}.


\section{Calculations}   \label{sec:calc}


We present calculations of the beam single spin asymmetry in a model where the nucleon is represented as a bound state of a spin-1/2 quark $q$ and a scalar diquark $S$, and where the momentum of the quark in the final state is measured.  Physically, the latter means that the quark jet in the final state is observed, and one measures its total momentum, but not any detailed  features that would give information about, for example, the polarization of the quark.  The fundamentals of the model are the same as in Ref.~\cite{bhs}.

The process, with momenta and helicities shown, is
\be
e(l,h_e) + N(p,\lambda) \rightarrow e(l',h_e) + q(p_1,\lambda') + S(p_2)     \,.
\ee
We treat the electron as massless, so that electron helicity $h_e = \pm 1/2$ is conserved.  The masses of the nucleon, quark, and scalar diquark will be $M$, $m$, and $m_S$, respectively.  The general matrix element is
\ba
{\cal M} &=& \frac{4\pi \alpha}{Q^2}    \langle e(l',h_e) | j_\nu | e(l,h_e) \rangle
				\nonumber \\
&&	\qquad 	\times \ 	\langle q(p_1,\lambda') \, S(p_2) | J^\nu | N(p,\lambda) \rangle \,,
\ea
where $j_\nu$ and $J^\nu$ represent electromagnetic current operators for electrons and hadrons, respectively, and the last matrix element will often be abbreviated $J^\nu(\lambda,\lambda')$.

The five-fold differential cross section for a polarized electron but unpolarized hadron is
\be
{d \sigma \over \left(dE_e'\, d\Omega'_e \right)^{\rm lab}  d\Omega_q^{\rm CM} } 
	= {\alpha^2 \over Q^4} 
	{E'_e \over E_e}
	{p_q^{\rm CM} \over 128 \pi^3 M W}  L_{\mu\nu}(h_e) W^{\mu\nu}   \,,
\ee
where $E_e$ and $E'_e$ are the  electron energies in the target rest frame (``lab''), $q=l-l^\prime$, 
$Q^2 = -q^2$,
$W^2 =(p+q)^2$, and $p_q^{\rm CM}$ is the quark three-momentum magnitude in the hadronic center-of-mass.  The lepton tensor is 
\ba
L_{\mu\nu}(h_e)  &=&  	\langle e(l,h_e) | j_\mu | e(l',h_e) \rangle
					\langle e(l',h_e) | j_\nu | e(l,h_e) \rangle
							\nonumber \\
			&=&		L^S_{\mu\nu} + (2h_e) L^A_{\mu\nu}    \,,
\ea
where ``\kern -1pt$S$'' or ``\kern -2pt$A$'' indicate terms that are symmetric or antisymmetric in the indices $(\mu,\nu)$.  The hadron tensor is here just
\be
W^{\mu\nu} = \frac{1}{2}  \sum_{\lambda,\lambda'}  \left( J^\mu(\lambda,\lambda') \right)^* 
				J^\nu(\lambda,\lambda')   \,.
\ee

Experimenters measure a single spin asymmetry defined from
\begin{equation}			\label{eq:asym}
d\sigma \propto \left( 1 + (2 h_e) A^{\sin\phi}_{LU} \sin\phi \right)      \,,
\end{equation}
whence
\be
A^{\sin\phi}_{LU} \sin\phi  =  \frac { L_{\mu\nu}^A W^{\mu\nu} }
		{ \langle  L_{\mu\nu}^S W^{\mu\nu} \rangle }  \,.
\ee
The pointed brackets indicate an averaging over the quark azimuthal angle $\phi$.  (Such an averaging was also done for the unpolarized term in Eq. (\ref{eq:asym}).)

After some modest effort, the last equation can be converted to
\be					\label{eq:altasym}
A^{\sin\phi}_{LU} \sin\phi  = - \frac { \sqrt{2\epsilon(1-\epsilon)} \, {\rm Im\,} W^{L2} }
		{ \epsilon W_{LL} +\frac{1}{2} \left( W_{11} + W_{22} \right) }  \,.
\ee
The indices ``$L$'' on the hadronic tensor indicate contraction with the exchanged photon's longitudinal polarization vector $\epsilon_L$,
\be
W^{L2} = W^{Ly} =  (\epsilon_L)_\mu W^{\mu 2} \,, \quad
	W_{LL} = (\epsilon_L)_\mu W^{\mu\nu} (\epsilon_L)_\nu  \,.
\ee
Plain $\epsilon$ is the photon polarization parameter, given by
\begin{equation}
{1\over \epsilon} = 1 + 2\left( 1 + {\nu^2 \over Q^2} \right)
			\tan^2 {\theta_e \over 2} \ ,
\end{equation}

\noindent where $\nu = E_e-E'_e$ and $\theta_e$ is the electron scattering angle in the target rest frame.  An alternative expression for $\epsilon$ using
\begin{equation}
y = \frac{\nu}{E_e}  \quad {\rm and} \quad \tau = \frac{\nu^2}{Q^2}
\end{equation}
is
\begin{equation}
\epsilon = \frac{1-y - y^2/(4\tau)}{1-y+y^2/2 + y^2/(4\tau)} \,.
\end{equation}
In the scaling limit ($\nu\to\infty$ with $y$ fixed), $\tau \to \infty$.

For spin-1/2 quarks, the transverse amplitudes dominate the longitudinal ones by a power of $Q$ in the scaling limit, and $\epsilon$ goes to a constant in the same limit.  Hence the transverse terms are dominant in the denominator of Eq.~(\ref{eq:altasym}) above, and the asymmetry $A^{\sin\phi}_{LU}$ falls like $1/Q$ at high $Q$.

We need to make some comments about gauge invariance.  Such discussion was unnecessary in~\cite{bhs}.  Those authors calculated a target spin asymmetry which depends only on amplitudes  with transversely polarized photons, which automatically conserve the electromagnetic current.  Beam spin asymmetry depends on an interference between longitudinal and transverse photon amplitudes, and longitudinal amplitudes are gauge sensitive.

\begin{figure}[h]

\medskip

\includegraphics[width=3.3in]{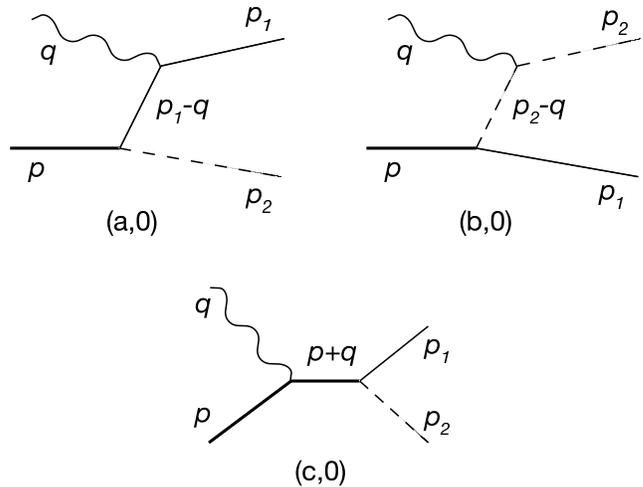}

\caption{Lowest order graphs.}

\label{fig:lowest}

\end{figure}

Gauge invariance requires that we allow the photon to interact with all charged particles in the Feynman diagram.  For $\gamma^* + N \to q + S$, this gives in lowest order the diagrams in Fig.~\ref{fig:lowest}.   There is a ``quark graph,'' a ``diquark graph,'' and a ``proton pole graph.''  In~\cite{bhs}, only the quark graph [(a,0)] was included.  In the large $Q$ limit and for transverse photons, the other two graphs are smaller by at least a factor $1/Q$.  For longitudinal photons, the additional graphs are not suppressed.


In general, one has all three diagrams.  Charge conservation, $e_N = e_1 + e_2$, ensures gauge invariance.  (Charges  $e_N$, $e_1$, and $e_2$, are for the nucleon, quark, and diquark, respectively.)

We calculate in a Breit frame, and find it convenient to adopt much of the notation of Ref.~\cite{Metz:2004je}.  We use the notation $\Delta_\perp$ for the outgoing quark transverse momentum.  The external momenta, in light front coordinates ($q^\pm = q^0 \pm q^3$), are
\ba
q &=& \left( q^+, q^-, q_\perp \right) =  \left( -Q, Q, 0_\perp \right)  \,,  \nonumber \\
p &=& \left( \frac{Q}{x} , \frac{xM^2}{Q} , 0_\perp \right)  \,, \nonumber \\
p_1 &=& \left( \frac{\Delta_\perp^2 + m^2}{Q} , 
	  Q + \frac{x}{Q} \Big( M^2 - \frac{\Delta_\perp^2 + m_S^2}{1-x} \Big) ,
	  \vec \Delta_\perp \right)  \,,
	  								\nonumber \\
p_2 &=& \left( Q \frac{1-x}{x} - \frac{\Delta_\perp^2 + m^2}{Q} , 
	  		\frac{x}{Q} \frac{\Delta_\perp^2 + m_S^2}{1-x} ,
	  		- \vec \Delta_\perp \right)  .
\ea
The electron scattering plane defines the $\hat x$-$\hat z$ plane, with the outgoing electron having a positive component in the $\hat x$ direction.  Momenta $p$ and $q$ are exact.  In the high $Q^2$ limit $x$ becomes the Bj\"ork\'en variable [$x_{Bj} = Q^2/ 2p\cdot q$]; momenta $p_{1,2}$ are accurate to ${\cal O}(1/Q)$.

In the Breit frame and using gauge invariance, one can show
\be
\left( \epsilon_L \right)_\mu J^\mu = J^+  \,.
\ee

The zero-loop diagrams give an amplitude $J^\mu_{(0)}(\lambda,\lambda')$ which is the sum of
\ba
J^\mu_{(a,0)} &=&  \frac{-e_1 g}{ (p_1-q)^2-m^2} \, \bar u(p_1,\lambda')
	\gamma^\mu ( \not\! p_1 - \not\! q + m) u(p,\lambda)   \,,
				\nonumber	\\			\nonumber 
J^\mu_{(b,0)} &=& - \frac{e_2 g}{ (p_2-q)^2 - m_S^2} \, 
	\left( 2p_2^\mu - q^\mu \right)
	\bar u(p_1,\lambda')   u(p,\lambda)   \,,
					\\			 \nonumber 
J^\mu_{(c,0)} &=& - \frac{e_N g}{ W^2 - M^2} \, 
	\bar u(p_1,\lambda')
	\left( \not\! p + \not\! q + M \right) \gamma^\mu u(p,\lambda)   \,.  \\
\ea

\noindent Thus
\begin{widetext}
\ba
J^2_{(a,0)}(\lambda,\lambda') &=& 
		\frac{e_1 g}{ \tilde m^2 + \vec\Delta_\perp^2 }
		\frac{1-x}{\sqrt{x} }
		\frac{Q}{ \sqrt{m^2 + \vec\Delta_\perp^2} }
	\bigg\{ i\lambda \delta_{\lambda,\lambda'}
			\Big[ (m+x M)  | \vec \Delta_\perp | 
				e^{-i\lambda \phi}
			- m | \vec \Delta_\perp | e^{i\lambda \phi}
			\Big]
					\nonumber \\
&& \hskip 12.5em  + \	i e^{i\lambda \phi} \delta_{\lambda,-\lambda'}
			\Big[ m(m+x M)  e^{-i\lambda \phi}
			+ | \vec \Delta_\perp |^2 e^{i\lambda \phi}
			\Big]
	\bigg\}            \,,
												\\
J^+_{(a,0)}(\lambda,\lambda') &=& 
		\frac{2 e_1 g}{ \tilde m^2 + \vec\Delta_\perp^2 }
		\frac{1-x}{\sqrt{x} }
		\sqrt{m^2 + \vec\Delta_\perp^2}
\  \ 		\bigg\{ \delta_{\lambda,\lambda'}
			(m+x M)
-		\lambda \delta_{\lambda,-\lambda'}
			| \vec \Delta_\perp | e^{i\lambda \phi}
	\bigg\}                  \,,
												\\
J^+_{(b,0)}(\lambda,\lambda') &=& 
		{e_2 g} 
		\frac{2-x}{\sqrt{x} }
		\frac{1}{ \sqrt{m^2 + \vec\Delta_\perp^2} }
\ \ 		\bigg\{  m \delta_{\lambda,\lambda'}
	-		\lambda \delta_{\lambda,-\lambda'}
				| \vec \Delta_\perp | e^{i\lambda \phi}
	\bigg\}  \,,
												\\
J^+_{(c,0)}(\lambda,\lambda') &=& 
		{ e_N g} 
		\frac{2}{\sqrt{x} }
		\frac{1}{ \sqrt{m^2 + \vec\Delta_\perp^2} }
\ \ 		\bigg\{ -  m \delta_{\lambda,\lambda'}
+		\lambda \delta_{\lambda,-\lambda'}
			| \vec \Delta_\perp | e^{i\lambda \phi}
	\bigg\}    \,.
\ea

\end{widetext}

\noindent
where,
\begin{equation}							\label{mtilde}
\tilde m^2 = x(1-x) \left( -M^2 + \frac{m^2}{x} + \frac{m_s^2}{1-x} \right)
\,.
\end{equation}

\noindent
Matrix elements $J^2_{(b,0)}$ and $J^2_{(c,0)}$ are subleading in $1/Q$.   Regarding $J^1_{(a,0)}$, we only need to know that $|J^1_{(a,0)}(\lambda,\lambda')|=|J^2_{(a,0)}(\lambda,\lambda')|$.

\begin{figure}[t]

\includegraphics[width=3.3in]{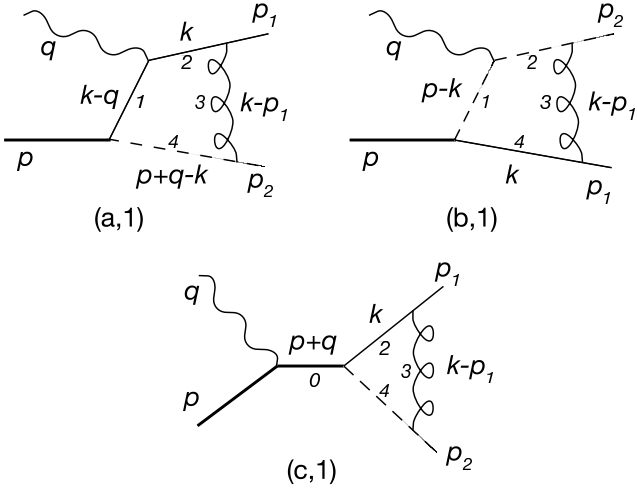}

\caption{One-loop graphs that have imaginary parts.}

\label{fig:oneloop}

\end{figure}

The calculated beam spin asymmetry, Eq.~(\ref{eq:altasym}), is zero if we have only the lowest order amplitudes.   To obtain a non-zero beam spin asymmetry we need to include next-to-leading order amplitudes and obtain a non-zero phase relative to the lowest order terms.  The diagrams are shown in Fig~\ref{fig:oneloop}.  Only one-loop diagrams that give an imaginary part are shown.

In the model, the final state interaction is mediated by a gluon that couples to a strong charge which is carried by the quark (strong charge  $e_s$) and diquark (strong charge  $(-e_s)$) but not by the proton.  One obtains the QCD equivalent by letting $e_s^2 \to C_F (4\pi\alpha_s)$ with $C_F=4/3$.  The diagrams give
\begin{widetext}

\ba					\label{eq:loops}
J^\mu_{(a,1)} &=& -i e_1 g e_s^2 \int \frac{d^4k}{16\pi^4}
	\frac{
	\bar u(p_1,\lambda') \left( \not\! p_1 + 2\!\not\! p_2 - \!\not\! k \right)
	\left( \not\! k + m \right) \gamma^\mu 
	\left( \not\! k - \!\not\! q +m \right) u(p,\lambda) }
	{ \left[ k^2 -m^2 + i\epsilon \right] 
		\left[ (k-q)^2 - m^2 + i\epsilon \right]
		\left[ (p+q-k)^2 - m_S^2 + i\epsilon \right]
		\left[ (k-p_1)^2 - \mu^2 + i\epsilon \right]} \,,
								\nonumber \\
J^\mu_{(b,1)} &=& -i e_2 g e_s^2 \int \frac{d^4k}{16\pi^4}
	\frac{
	\bar u(p_1,\lambda') \left( \not\! p_1 + 2\!\not\! p_2 - \!\not\! k \right)
	\left( \not\! k + m \right) u(p,\lambda)  
	\left(2p^\mu+q^\mu-2k^\mu\right) }
	{ \left[ k^2 -m^2 + i\epsilon \right] 
		\left[ (p-k)^2 - m_S^2 + i\epsilon \right]
		\left[ (p+q-k)^2 - m_S^2 + i\epsilon \right]
		\left[ (k-p_1)^2 - \mu^2 + i\epsilon \right]} \,,
								\nonumber \\
J^\mu_{(c,1)} &=& -i e_N g e_s^2 \int \frac{d^4k}{16\pi^4}
	\frac{
	\bar u(p_1,\lambda') \left( \not\! p_1 + 2\!\not\! p_2 - \!\not\! k \right)
	\left( \not\! k + m \right) 
	\left( \not\! p + \!\not\! q + M \right) \gamma^\mu  u(p,\lambda) }
	{ \left[ k^2 -m^2 + i\epsilon \right] 
		\left[ (p+q)^2 - M^2 + i\epsilon \right]
		\left[ (p+q-k)^2 - m_S^2 + i\epsilon \right]
		\left[ (k-p_1)^2 - \mu^2 + i\epsilon \right]} \,,
\ea

\noindent
where $\mu$ in the denominator is a small gluon mass temporarily included.  While present in the amplitudes, it will not appear in the final answer for beam spin asymmetry $A_{LU}$ indicating that the considered observable is
infrared-safe.
The absorptive part (the part that is imaginary relative to the lowest order) is obtained by substituting
\ba					\label{eq:abs}
\frac{1}{ k^2 -m^2 + i\epsilon } \frac{1}{ (p+q-k)^2 - m_S^2 + i\epsilon }
	\stackrel{\to}{=} 2i\pi^2 
		\delta_+( k^2 -m^2 ) \delta_+( (p+q-k)^2 - m_S^2 )  \,,
\ea

\noindent

As a result, the 4--dimensional integration over the loop momentum in Eq. (\ref{eq:loops}) is reduced to a 2--dimensional angular integration, making the result also safe from ultraviolet divergence.
In the numerical calculations, we found projections of the electromagnetic current Eq. (\ref{eq:loops})
on six independent helicity amplitudes of the process $\gamma^*+N\to q+S$, and $A_{LU}$ was expressed in terms of these helicity amplitudes. Two-dimensional integration with respect to polar and azimuthal
angles of the intermediate quark was done numerically with {\it Mathematica} in 
the $\gamma^*+N$ center-of-mass frame.
Independence of $A_{LU}$ on the gluon cut-off mass $\mu$  was verified, and  
providing  a cross-check of the numerical calculation.  
The plots of  the asymmetry $A_{LU}$ shown later in this paper were obtained using the procedure
described in this paragraph.

We may also obtain approximate analytic formulas for the beam spin asymmetries in the limit where the masses and transverse jet momentum are much smaller than the invariant momentum transfer $Q$, by keeping leading terms in a $1/Q$ expansion.  To begin, we have
\ba
{\rm Abs\,} J^\mu_{(a,1)}(\lambda,\lambda') &=& \frac{e_1 g e_s^2}{8\pi^2}
	\frac{x}{Q^4}  \int d^2 k_\perp 
	\frac{ 
	\bar u(p_1,\lambda') \!\not\! p_2 \left( \not\! k + m \right) 
	\gamma^\mu  \left( \not\! k - \!\not\! q +m \right)
	u(p,\lambda)  }
	{  \left( \vec k_\perp^2 + \tilde m^2 \right)
	\left( \big( \vec k_\perp - \Delta_\perp \big)^2 + \mu^2 \right) }  \,,
								\nonumber \\
{\rm Abs\,} J^\mu_{(b,1)}(\lambda,\lambda') &=& \frac{e_2 g e_s^2}{8\pi^2}
	\frac{x^2}{1-x} \frac{1}{Q^4}  \int d^2 k_\perp 
	\frac{ 
	\bar u(p_1,\lambda') \!\not\! p_2 \left( \not\! k + m \right) u(p,\lambda)  
	\left( 2p^\mu+q^\mu-2k^\mu \right) }
	{  \left( \big( \vec k_\perp - \Delta_\perp \big)^2 + \mu^2 \right) } \,,
								\nonumber \\
{\rm Abs\,} J^\mu_{(c,1)}(\lambda,\lambda') &=& - \frac{e_N g e_s^2}{8\pi^2}
	\frac{x^2}{ (1-x)^2 } \frac{1}{Q^4}  \int d^2 k_\perp 
	\frac{ 
	\bar u(p_1,\lambda') \!\not\! p_2 \left( \not\! k + m \right) 
	\left( \not\! p + \!\not\! q + M \right) \gamma^\mu
	u(p,\lambda)  }
	{ \left( \big( \vec k_\perp - \Delta_\perp \big)^2 + \mu^2 \right) }  \,.
\ea

The $k_\perp$ integrals have a finite upper limit because the quark and diquark in the intermediate state are now on-shell.  Doing the integrals yields 
\ba
&&{\rm Abs\,} J^2_{(a,1)}(\lambda,\lambda') =
	\frac{e_s^2}{8\pi}  J^2_{(a,0)}(\lambda,\lambda')  L_{\Delta\mu}
+	\frac{e_1 g e_s^2}{8\pi \big(\tilde m^2 + \vec\Delta_\perp^2 \big)}
	\frac{1-x}{\sqrt{x} }
	\frac{Q}{ \sqrt{m^2 + \vec\Delta_\perp^2} }  \times
							\nonumber \\
&&\times\,	\Bigg\{ i\lambda \delta_{\lambda,\lambda'}
	\left[  (m+xM) |\vec \Delta_\perp | e^{-i\lambda \phi}
		+ \frac{ m \tilde m^2  }
			{ |\vec \Delta_\perp | } e^{i\lambda \phi} 
	\right]
	+	ie^{i\lambda \phi} \delta_{\lambda,-\lambda'}
	\bigg[   m(m+xM) e^{-i\lambda \phi}
		- \tilde m^2  e^{i\lambda \phi} 
	\bigg]
	\Bigg\}   L_{\Delta m}	 \,,
							\\
&&{\rm Abs\,} J^+_{(a,1)}(\lambda,\lambda') =
	\frac{e_s^2}{8\pi}  J^+_{(a,0)}(\lambda,\lambda') L_{\Delta\mu}
+	\frac{2 e_1 g e_s^2}{8\pi (\tilde m^2 + \vec\Delta_\perp^2) }
	\frac{1-x}{\sqrt{x} }
	\frac{1}{ \sqrt{m^2 + \vec\Delta_\perp^2} }  \times
							\nonumber \\
&&\times\,	\Bigg\{ \delta_{\lambda,\lambda'}
	\left[   (m+xM) 	 \left(   (m^2+ \vec \Delta_\perp^2 ) - (\tilde m^2 + \vec\Delta_\perp^2)  
				 \right)  L_{\Delta m}	 
		+ m (\tilde m^2 + \vec\Delta_\perp^2)  \left( L_{Q\Delta} - 1 \right) 
		\right]
							\nonumber \\
&&\quad +\	\lambda \delta_{\lambda,-\lambda'}
	\left[    \left(  - (m^2+ \vec \Delta_\perp^2 ) 
			+ \left( 1 - \frac{xmM}{|\vec\Delta_\perp|^2} \right) 
			( \tilde m^2+ \vec \Delta_\perp^2 )  \right) 
			L_{\Delta m}	
		- ( \tilde m^2+ \vec \Delta_\perp^2 )  ( L_{Q\Delta} - 1 ) 
	\right]
	|\vec \Delta_\perp |  e^{i\lambda \phi}
	\Bigg\}    \,,
							\\
&&{\rm Abs\,} J^+_{(b,1)}(\lambda,\lambda') =
	\frac{ e_s^2}{8\pi}
	J^+_{(b,0)}(\lambda,\lambda') 
	\bigg\{ 
	 L_{\Delta\mu} + L_{Q\Delta}  -  \frac{2(1-x)}{2-x}	\bigg\}    \,,
							\\
&&{\rm Abs\,} J^+_{(c,1)}(\lambda,\lambda') =
		\frac{ e_s^2}{8\pi}
	J^+_{(c,0)}(\lambda,\lambda') 
	\bigg\{ 
	 L_{\Delta\mu} + L_{Q\Delta}  - 1 \bigg\}    \,.
\ea

\noindent   or

\ba
&& {\rm Abs\,} J^+_{(1)} = {\rm Abs\,} J^+_{(a,1)} + {\rm Abs\,} J^+_{(b,1)}+ {\rm Abs\,} J^+_{(c,1)}
	= \frac{e_s^2}{8\pi}  J^+_{(0)}(\lambda,\lambda')  \, L_{\Delta\mu}
	+	\frac{ g e_s^2}{8\pi (\tilde m^2 + \vec\Delta_\perp^2) }
	\frac{1-x}{\sqrt{x} }
	\frac{1}{ \sqrt{m^2 + \vec\Delta_\perp^2} }  \times
							\nonumber \\
&&\times\,	\Bigg\{ \delta_{\lambda,\lambda'}
	\bigg[   2e_1 (m+xM) 	 \left(   m^2 - \tilde m^2    \right)  L_{\Delta m} 
	- \frac{x}{1-x} \Big(  \left(  2e_1 + e_2 \right)   L_{Q\Delta} - 2 \left( e_1 + e_2  \right) \Big)
		 m (\tilde m^2 + \vec\Delta_\perp^2) 
		 \bigg]
							\\ \nonumber 
&&\  	+\,	\lambda \delta_{\lambda,-\lambda'}
	\bigg[   2e_1   \bigg(  \tilde m^2 - m^2  - xmM 
		\frac{  \tilde m^2+ \vec \Delta_\perp^2 }{|\vec\Delta_\perp|^2}  \bigg) 
			L_{\Delta m}
	+ \frac{x}{1-x} \Big(  \left(  2e_1 + e_2 \right)   L_{Q\Delta} - 2 \left( e_1 + e_2  \right) \Big) 
		( \tilde m^2+ \vec \Delta_\perp^2 )
			\bigg]
	|\vec \Delta_\perp |  e^{i\lambda \phi}
	\Bigg\}    \,,
\ea
using charge conservation.

\noindent
Here
\ba
L_{\Delta m}&=&  \ln\frac{ \tilde m^2 + \vec\Delta_\perp^2 }{\tilde m^2}  \,, \qquad
L_{\Delta\mu} =  \ln\frac{ \tilde m^2 + \vec\Delta_\perp^2 }{\mu^2}  \,, \qquad
L_{Q\Delta} =  \ln\frac{Q^2 (1-x)}{ (\tilde m^2 + \vec\Delta_\perp^2) x}  \,.
\ea

The general result for the beam spin asymmetry is
\begin{eqnarray}   \label{general_alu}
A_{LU}^{\sin\phi} &=& \frac{e_s^2}{4\pi} \sqrt{2\epsilon(1-\epsilon)}
	\frac{ \tilde m^2 + \vec\Delta_\perp^2 }{(m+Mx)^2 + \vec\Delta_\perp^2}
		\frac{ |\vec\Delta_\perp| }{Q}
				\\ \nonumber
	&\times&   \Bigg\{ \frac{1}{\vec\Delta_\perp^2}
		\left( M^2x^2 - m^2 + \frac{2e_1-xe_2}{2e_1(1-x)} \, \tilde m^2 \right)
			\ln\frac{ \tilde m^2 + \vec\Delta_\perp^2 }{ \tilde m^2 }
	+  \frac{2e_1+e_2}{2e_1}	\frac{x}{1-x} \ln
		\frac{Q^2 (1-x)}{ \left( \tilde m^2 + \vec\Delta_\perp^2 \right) x }
			- \frac{e_1+e_2}{e_1}		\frac{x}{1-x}
	\bigg\}      \,.
\end{eqnarray}

\end{widetext}


The asymmetry falls like $1/Q$, as expected for a twist-3 observable.

For numerical work we need to make choices for the charges, and we will show results for several possibilities.  Two choices, will be for up-quark and down-quark jets from a proton, where $e_1$ is the charge for the stated quark, $e_N = 1$, and $e_2$ is the remainder.  Another proton-like choice has just the nucleon and quark diagrams, with $e_N = e_1$ and with the diquark electrically neutral.  The explicit asymmetry for this case is,
\begin{eqnarray}     \label{alu_p}
A_{LU}^{\sin\phi} &=& \frac{e_s^2}{4\pi} \sqrt{2\epsilon(1-\epsilon)}
	\frac{ \tilde m^2 + \vec\Delta_\perp^2 }{(m+Mx)^2 + \vec\Delta_\perp^2}
		\frac{ |\vec\Delta_\perp| }{Q}
				\nonumber \\
	\times&& \hskip -14pt \Bigg\{ \frac{1}{\vec\Delta_\perp^2}
		\left( M^2x^2 - m^2 + \frac{1}{1-x} \, \tilde m^2 \right)
			\ln\frac{ \tilde m^2 + \vec\Delta_\perp^2 }{ \tilde m^2 }
				\nonumber \\
	&+& \frac{x}{1-x} \ln
		\frac{Q^2 (1-x)}{ \left( \tilde m^2 + \vec\Delta_\perp^2 \right) x }
			- \frac{x}{1-x}
	\Bigg\} \,.
\end{eqnarray}

A further case keeps just the quark and diquark diagrams.  This case is neutron-like, since the nucleon is chosen electrically neutral, and $e_2=-e_1$.  This is the prescription used by Metz and Schlegel~\cite{Metz:2004je} to ensure gauge invariance, and it gives
\begin{eqnarray}       						\label{alu_n}
A_{LU}^{\sin\phi} &=& \frac{e_s^2}{4\pi} \sqrt{2\epsilon(1-\epsilon)}
	\frac{ \tilde m^2 + \vec\Delta_\perp^2 }{(m+Mx)^2 + \vec\Delta_\perp^2}
		\frac{ |\vec\Delta_\perp| }{Q}
				\nonumber \\
	\times&& \hskip -14pt \Bigg\{ \frac{1}{\vec\Delta_\perp^2}
		\left( M^2x^2 - m^2 + \frac{2-x}{2(1-x)} \,\tilde m^2 \right)
			\ln\frac{ \tilde m^2 + \vec\Delta_\perp^2 }{ \tilde m^2 }
				\nonumber \\
	&+& \frac{x}{2(1-x)} \ln
		\frac{Q^2 (1-x)}{ \left( \tilde m^2 + \vec\Delta_\perp^2 \right) x }
	\Bigg\}  \,.
\end{eqnarray}
The result can be checked against~\cite{Metz:2004je} in the limit $m\to 0$.

Finally, in our earlier note~\cite{Afanasev:2003ze}, we implemented gauge invariance using only the quark diagram [(a)], but modifing its amplitude $J^\mu$ by subtracting out the scalar part~\cite{Gross:1987bu},
\be
\label{eq:subtraction}
J^\mu \to J^\mu - q^\mu \frac{q\cdot J}{q^2} \,.
\ee
For comparison, we report the result from this ``$q^\mu$ subtraction'' method here.  It is
\begin{eqnarray}							\label{alu_qmu}
A_{LU}^{\sin\phi} &=& \frac{e_s^2}{4\pi} \sqrt{2\epsilon(1-\epsilon)}
	\frac{ \tilde m^2 + \vec\Delta_\perp^2 }{(m+Mx)^2 + \vec\Delta_\perp^2}
		\frac{ |\vec\Delta_\perp| }{Q}
				\nonumber \\
	\times&& \hskip -14pt \Bigg\{ \frac{1}{\vec\Delta_\perp^2}
		\left( M^2x^2 - m^2 + \frac{1}{2(1-x)} \,\tilde m^2 \right)
			\ln\frac{ \tilde m^2 + \vec\Delta_\perp^2 }{ \tilde m^2 }
				\nonumber \\
	&-& \frac{1-2x}{2(1-x)} \ln
		\frac{Q^2 (1-x)}{ \left( \tilde m^2 + \vec\Delta_\perp^2 \right) x }
			+ \frac{ 1-2x }{ 2(1-x) }
	\Bigg\}  \,.
\end{eqnarray}


\begin{figure}
\vspace{-0.2in}
\includegraphics[width=3.3in]{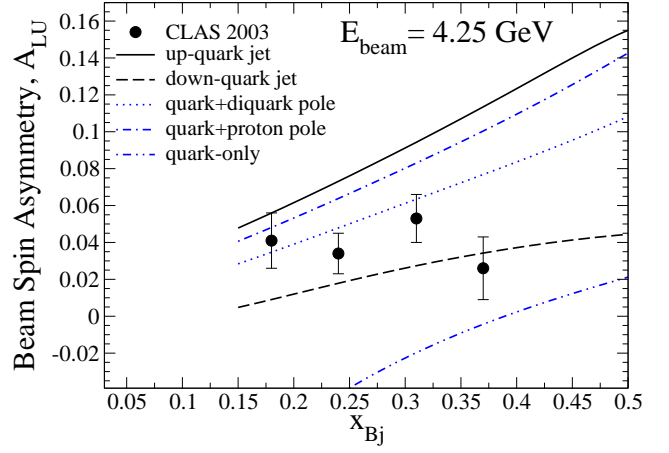}
\caption{Beam single spin asymmetry for quark jets of different flavors and beam energy of 4.25 GeV.
Data are from Ref.\cite{Avakian:2003pk}. See the text for notation.}
\label{fig:4.25gev}
\end{figure}



\begin{figure}[t]
\vspace{0.1in}

\includegraphics[width=3.3in]{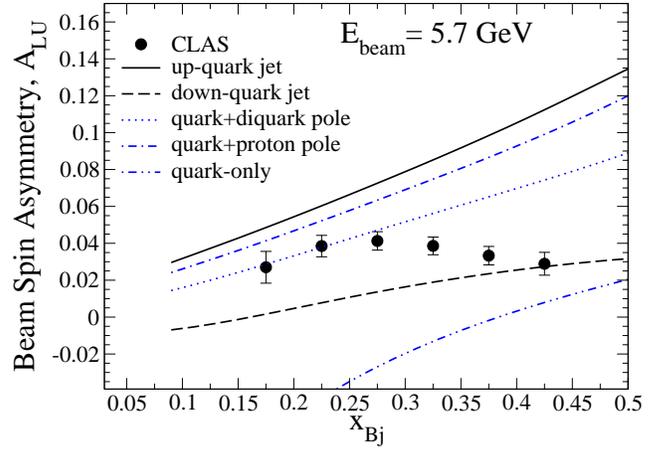}

\caption{Beam single spin asymmetry for quark jets of different flavors and beam energy of 5.7 GeV.
Data are from Ref.\cite{Avakian:2003pk}. Notation is as in Fig.\ref{fig:4.25gev}.}

\label{fig:5.7gev}

\end{figure}

\begin{figure}[h]
\vspace{0.25in}
\includegraphics[width=3.3in]{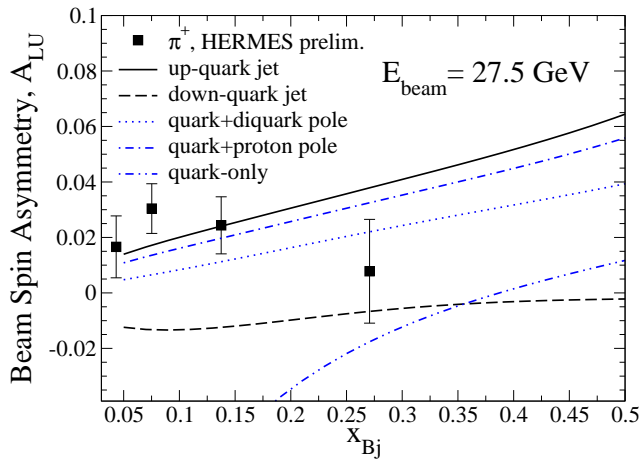}

\caption{Beam single spin asymmetry for quark jets of different flavors and beam energy of 27.5 GeV.
Also shown are preliminary data from HERMES \cite{hermes_ssa}. Notation is as in Fig.\ref{fig:4.25gev}.}

\label{fig:27.5gev}

\end{figure}

Numerical results are given in the form of plots of $A_{LU}^{\sin\phi}$ vs. $x_{bj}$ for three different beam energies.   For all the plots, we use parameter values $M = 939$ MeV, $m = 300$ MeV, $m_S = 800$ MeV, and $\alpha_s  = 0.3$.  The quark transverse momentum $|\vec \Delta_\perp|$ is fixed at 0.4 GeV.   Results for beam energies of 4.25 GeV, 5.7 GeV, and 27.5 GeV, for quark jets of different flavors and with different versions of the model, are shown in Figs.~\ref{fig:4.25gev},~\ref{fig:5.7gev}, and~\ref{fig:27.5gev}. The dashed-double-dotted curve (up to a corrected sign) matches  results of our earlier report \cite{Afanasev:2003ze}. For lower beam energies, we start the plots from larger values of $x_{Bj}$
to ensure the condition $Q^2>M^2$ holds.
The solid and dashed curves, respectively, describe the asymmetry in electroproduction of up- and down-quark jets if all three diagrams of Fig. 2 are included. If the diquark is assumed to be not interacting with
electromagnetic probe, only the diagrams Fig. 2(a) and (c) contribute, and the corresponding curve is  dashed-dotted. If the nucleon target is electrically neutral, then the virtual photon couples only
to the quark and an (oppositely-charged) diquark; the asymmetry appears equal for both quark jet flavors.
This situation is described by a dotted curve in Figs.~\ref{fig:4.25gev},~\ref{fig:5.7gev}, and~\ref{fig:27.5gev}. 

The results on these plots are from the numerically integrated results obtained as described just after Eq. (\ref{eq:abs}).  The analytic forms valid for large $Q$ are essentially identical to these results for the $E_e= 27.5$ GeV case, and are relatively 10--20\% lower for the other two energies.  

Remembering that our calculations refer to quark jets with no fragmentation
into hadrons, we can see that both the sign and the magnitude of the asymmetry $A_{LU}$ compare favorably
with experimental data from HERMES and JLab CLAS. We also note that the experiment observed somewhat stronger
supression of the asymmetry at larger $x_{Bj}$ compared to our model predictions.


\section{The distribution function $g^\perp$}   \label{sec:gperp}


The general expression for the beam spin asymmetry can be summarized by the transverse momentum dependent (TMD) distribution function $g^\perp(x,\vec\Delta_\perp^2)$, for the situation that only the outgoing quark jet is observed and in the limit of neglecting the quark mass.  Until relatively recently, for the quark jet case,  the known lists of initial state quark distribution functions contained none  that could give a beam spin asymmetry, although beam spin asymmetry was known in cases that involved quark fragmentation functions that gave information about quark spin correlations. Now, with the work of Yuan\cite{Yuan:2003gu},  Bacchetta {\it et al.}~\cite{Bacchetta:2004zf}, and Goeke {\it et al.}~\cite{Goeke:2005hb},  the distribution functions $h_1^\perp(x,p_T^2)$ and $g^\perp(x,p_T^2)$ have been uncovered, and they allow an observable beam spin asymmetry without any real or implied spin measurement in the final state.   The asymmetry due to $h_1^\perp$ is proportional to the quark mass, and we consider in this Section only the massless quark case, and for simplicity also let $m_S = M$.

From Eqs. (15--17) of~\cite{Bacchetta:2004zf}, suitably tuned for the present case and using the sign convention from~\cite{Goeke:2005hb} (that is opposite to \cite{Bacchetta:2004zf}), one obtains
\be
A_{LU}^{\sin\phi} = \sqrt{2\epsilon(1-\epsilon)} \frac{|\vec\Delta_\perp|}{Q}
	\frac{ x g^\perp(x,\vec\Delta_\perp^2)}{f_1(x,\vec\Delta_\perp^2)}  \,.
\ee

\noindent
The TMD function $f_1(x,\vec\Delta_\perp^2)$ gives the probability distribution of quarks in a hadron, and is related to the hadron matrix by
\be
W^{11} = W^{22} = 2(2\pi)^3 2M\nu f_1(x,\vec\Delta_\perp^2)  \,.
\ee

\noindent
Thus, in the present model, 
\be
 f_1(x,\vec\Delta_\perp^2)
		= \frac{1}{16\pi^3}  \frac{ g^2 (1-x)^2 }{ ( \tilde m^2 + \vec\Delta_\perp^2 )^2 }
			\left\{  (m+xM)^2  +  \vec \Delta_\perp  ^2  \right\} ,
\ee

\noindent  (before $m$ is set to zero).  We obtain a value for $g$ from the normalization condition,
\be
N_q = 1 = \int_0^1 dx 
	\int_{\vec\Delta_\perp^2 = 0}^{\vec\Delta_\perp^2 = \Lambda^2} d^2\Delta_\perp  \, 
		f_1(x,\vec\Delta_\perp^2)     \,.
\ee
The distribution $f_1$ is calculated from lowest order diagrams, so that the momentum $\vec\Delta_\perp$ is identical to the transverse momentum of the quark in the initial state.  We pick a cutoff pertinent to the process at hand, in particular we choose it equal to the maximum value of transverse momentum for a real quark in the final state,
\be
\Lambda^2 = \left.  \vec\Delta_\perp^2  \right|_{max}  =  Q^2 \frac{1-x}{4x}  \,.
\ee
Thus,
\be
\frac{g^2}{16 \pi^2} = \left[ \frac{1}{3} \ln \left( \frac{Q^2}{4 M^2} \right)  + \frac{31}{18} \right]^{-1}  \,.
\ee

\begin{widetext}
The result for $g^\perp$ is
\ba										\label{gperp}
g^\perp(x, \vec\Delta_\perp^2) &=& \frac{g^2}{16\pi^3}
	\frac{C_F \alpha_s   (1-x)^2 }{ x ( \tilde m^2 + \vec\Delta_\perp^2 ) }
	\\ \nonumber
	&\times&
		 \Bigg\{ \frac{1}{ \vec\Delta_\perp^2}
		\left( M^2x^2  + \frac{2e_1-xe_2}{2e_1(1-x)} \, \tilde m^2 \right)
			\ln\frac{ \tilde m^2 + \vec\Delta_\perp^2 }{ \tilde m^2 }
	+  \frac{2e_1+e_2}{2e_1}	\frac{x}{1-x} \ln
		\frac{Q^2 (1-x)}{ \left( \tilde m^2 + \vec\Delta_\perp^2 \right) x }
			- \frac{e_1+e_2}{e_1}		\frac{x}{1-x}
	\bigg\}   \,.
\ea

\end{widetext}

\noindent The quantity in curly brackets is the same as Eq.~(\ref{general_alu}), except that the mass $m$ is set to zero both there and in Eq.~(\ref{mtilde}).  If desired, one may also substitute the contents of the curly brackets for the special cases shown in Eqs.~(\ref{alu_p}), (\ref{alu_n}), or (\ref{alu_qmu}).

 Again, our calculation of the beam single spin asymmetry does not depend on factorization. Obtaining $g^\perp$ does depend on the analyses of Refs.~\cite{Bacchetta:2004zf} and~\cite{Goeke:2005hb}, and hence does assume factorization.  Factorization appears to have been proved for some twist-2 transverse momentum dependent distributions with transversely polarized targets~\cite{Ji:2004wu,Collins:2004nx}.  For longitudinal polarization and for twist-3 distributions, the question of factorization has been raised far in the past~\cite{leveltmulders} and continues to be discussed ({\it e.g.},~\cite{Gamberg:2006ru}).  Factorization for these cases remains assumed rather than proved, and this should be borne in mind.

There is a $\log Q^2$ term in $g^\perp$, mirroring a similar term in the result for the beam spin asymmetry.  The possibility of such terms arises because the asymmetry and $g^\perp$ are non-zero only when loop corrections are included.  Loops, however, do not always lead to $\log Q^2$ terms;  it depends on the details of the integrals.  For contrast, there is  the twist-2 target spin asymmetry or Sivers effect, where the integrals are insensitive to their upper limits in a similar one-loop calculation~\cite{bhs} and consequently one finds no $\log Q^2$ term.  We interpret the $\log Q^2$ simply as $g^\perp$ possessing a logarithmic scaling violation within this model.  It is not, as far as we can see, related to questions regarding factorization. 

\begin{figure}[h]
\vspace{0.2in}
\includegraphics[width=3.3in]{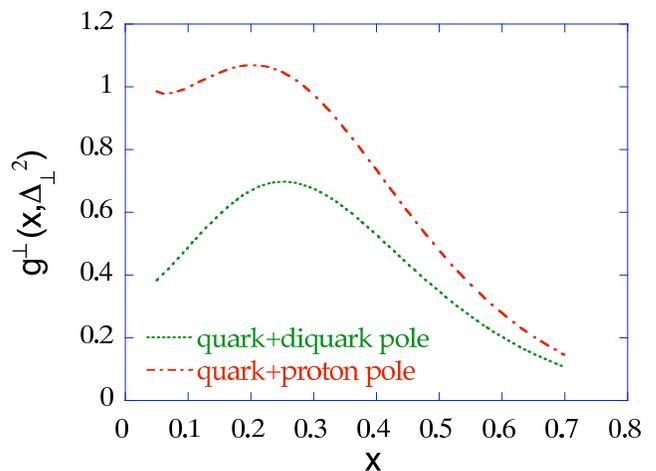}
\caption{The distribution function $g^\perp(x,\vec\Delta_\perp^2)$ for 
$| \vec\Delta_\perp | =0.4$ GeV.  The two special cases are described in detail in the text. A short summary is that the nucleon in the quark+diqurk pole case is overall electrically neutral, and in the quark+proton pole case has unit charge.  In both cases all flavors of quark in the final state are summed. }

\label{fig:gperp}

\end{figure}

A plot of $g^\perp(x,\vec\Delta_\perp^2)$ is shown  in Fig.~\ref{fig:gperp} for $| \vec\Delta_\perp | = 0.4$ GeV and varying $x$, for two of the special cases described earlier.  Except for the logarithmic $Q^2$ dependence induced in the normalization condition, the result is independent of beam energy.  (The range of $x$ for which the curve is plotted is fixed on the left by requiring $Q^2$ to be above $M^2$ and on the right by requiring that $W$, the final state hadronic mass, is above 2 GeV.  This figure as shown is for $E_{beam} = 27.5$ GeV.)


\section{Beam Spin Asymmetry for Scalar Quarks}   \label{sec:scalar}


The fact that there is a beam single-spin asymmetry for an outgoing jet does not depend on the spin of the target or of its constituents.  We emphasize that here by calculating $A_{LU}^{\sin\phi}$ for hypothetical scalar quark bound states.

Some of the details are different, however.  In particular, for scalars the longitudinal amplitude dominates the transverse one at high $Q$, in contrast to the spinor case.  Also, considering just the case that the diquark is neutral, only the quark pole diagrams are needed for the cross section and asymmetry calculation.  To wit, using notation from Fig.~\ref{fig:lowest}, we have
\ba
J^\mu_{(a,0)} &=& - \frac{e_1 g}{ (p_1-q)^2-m^2} \,
		\left( 2p_1 - q \right)^\mu		   \,,
				\nonumber	\\			
J^\mu_{(c,0)} &=& - \frac{e_N g}{ (p+q)^2 - M^2} \, 
		\left( 2p + q \right)^\mu   \,.  
\ea
Both graphs, with $e_N=e_1$, are needed to satisfy gauge invariance.  However, for the longitudinal amplitudes,
\ba
J^+_{(a,0)} &=&  \frac{e_1 g (1-x)}{ \tilde m^2+ \vec \Delta_\perp^2 } Q	   \,,
				\nonumber	\\			
J^+_{(c,0)} &=& - \frac{e_1 g (2-x) }{ (1-x) Q}    \,, 
\ea
so that the proton pole graph contribution is subleading at high $Q$.  Similarly for the transverse amplitudes,
\ba
\vec J_{(a,0)\perp} &=&  \frac{e_1 g (1-x)}{ \tilde m^2+ \vec \Delta_\perp^2 } \vec \Delta_\perp	   \,,
				\nonumber	\\			
\vec J_{(c,0)\perp} &=& \vec 0_\perp    \,.
\ea

After also calculating the relevant one-loop graphs, the beam single-spin asymmetry for semi-inclusive jet production is
\be
A_{LU}^{\sin\phi} = \frac{4\alpha_s}{3} 
	\frac{\sqrt{2\epsilon(1-\epsilon)} Q} {\epsilon Q^2 + 2 \vec \Delta_\perp^2 } \,
		\frac{  \tilde m^2 + \vec \Delta_\perp^2 } { |\vec \Delta_\perp | }
			\ln \left(  \frac{  \tilde m^2 + \vec \Delta_\perp^2 }{  \tilde m^2 }  \right) \,,
\ee
for $Q$ large compared to masses and to $|\vec \Delta_\perp | $.   For the curious, evaluating the above scalar formula numerically with the same values of masses and $|\vec \Delta_\perp | $ mentioned previously gives about the  level of the (presumably spin-1/2 quark) data in Fig.~\ref{fig:4.25gev} if we use a rather small $\alpha = 0.075$.


\section{Discussion}   \label{sec:end}


We have calculated, in a definite model, the beam spin asymmetry for semi-inclusive inelastic electron scattering.  Experimentally, beam spin asymmetry has been seen at JLab/CLAS and at Hermes.  Our particular calculation considered beam spin dependent asymmetry in jet production, where all the final-hadrons in the jet are summed and their phase space integrated.

Beam spin asymmetry means an asymmetry in the direction of an outgoing hadron relative to a direction set by the polarization of the incoming beam.  In the case at hand, define an observable by first defining a normal from the electron spin and the photon momentum, $\vec S_e \times \vec q$.  Representing the momentum of the jet by the momentum 
$p_q$ of the quark, the beam spin asymmetry observable is 
$\vec S_e \times \vec q \cdot \vec p_q$.  If one defines a coordinate system, this observable is proportional to $\sin \phi$, where $\phi$ is the azimuthal angle between the lepton scattering and jet production planes.  Having a beam spin asymmetry means that the expectation value, weighted by the cross section, of this observable is non-zero.  

As a useful detail, the electron spin direction is communicated to the hadrons through the photon, and if the electrons are polarized, the photon aquires a vector polarization $\vec S_\gamma$ which is in the electron scattering plane but not parallel to $\vec q$.  An alternative beam spin asymmetry observable is 
$\vec S_\gamma \times \vec q \cdot \vec p_q$.  The observable is odd under time reversal.  Further, this observable is reminiscent of the single spin asymmetry observable for a polarized proton target,
$\vec S_p \times \vec q \cdot \vec p_q$, which measures the Sivers function~\cite{bhs,Anselmino:2005ea} and which also was thought for some time to be absent or zero because of time reversal invariance.

The beam spin asymmetry is zero in a lowest order calculation, as expected for a $T$-odd observable.
It is not zero in higher orders.  Hadronic final state interactions lead to a relative phase between the longitudinal and transverse amplitudes, and interference between these two amplitudes gives the beam spin asymmetry.

The model in which we calculated the beam spin asymmetry is a simple one where the proton is represented as a bound state of a quark and a scalar diquark, and where the hadronic final state interaction is produced by a gluon exchange.  Gauge invariance requires that we include all diagrams where the photon couples to a charged hadron.  There are three diagrams in lowest order, and also three relevant diagrams at one-loop order.   Results for the beam spin asymmetry $A_{LU}^{\sin\phi}$ have been given in the body of the paper both numerically in plots and as formulas for the case when masses and transverse jet momentum are much smaller than the invariant momentum transfer $Q$.

Quark spin not crucial to obtaining a beam spin asymmetry.  The qualitative effect persists in an equivalent model with scalar quarks, although the numerical results are not the same.

In modern times, the expression for the beam spin asymmetry in jet production can in general be given in terms of the initial state transverse momentum dependent parton distribution called 
$g^\perp(x,k_\perp^2)$, where $k_\perp$ is the transverse momentum of the struck quark relative to the parent proton.  The model thus gives an explicit result for $g^\perp$, which is shown in Eq.~(\ref{gperp}).


\begin{acknowledgments}

This work was supported by the US Department of Energy
under contract DE-AC05-84ER40150 (A.V.A.) and by the National Science Foundation
under grants PHY-0245056 (C.E.C.), PHY-0114343 and PHY-0301841 (A.V.A.)

\end{acknowledgments}


\end{document}